# The Optical Gravitational Lensing Experiment.
## Variable Stars in Globular Clusters -I. Fields 5139A-C in Omega Centauri [1]

J. Kaluzny, M. Kubiak, M. Szymański & A. Udalski

Warsaw University Observatory, Al. Ujazdowskie 4, 00-478 Warsaw, Poland
e-mail: (jka,mk,msz,udalski)@sirius.astrouw.edu.pl

and

W. Krzemiński

Carnegie Observatories, Las Campanas Observatory, Casilla 601, La Serena, Chile
e-mail: wojtek@roses.ctio.noao.edu

and

M. Mateo

Department of Astronomy, University of Michigan, 821 Dennison Bldg., Ann Arbor, MI
48109-1090, USA
e-mail: mateo@astro.lsa.umich.edu

## ABSTRACT

Three fields covering the central part of the globular cluster $\omega$ Cen were surveyed in a search for variable stars. We present V-band light curves for 22 periodic variables: 9 SX Phe stars, 7 contact binaries, 5 detached or semi-detached binaries, and one spotted variable (FK Com or RS CVn type star). Only 2 of these variables were previously known. All SX Phe stars and all contact binaries from our sample belong to blue stragglers. Observed properties of these stars are consistent with their cluster membership. Of particular interest is detection of two well detached binaries with periods P=1.50 day and P=2.47 day. Further study of these two binaries can provide direct information about properties of turnoff stars in $\omega$ Cen. An uncomplete light curve of a Mira variable known as V2 was obtained. We present $V$ vs. $V - I$ color-magnitude diagrams for the monitored part of the cluster.

## 1. Introduction

The Optical Gravitational Lensing Experiment (OGLE) is a long term project with the main goal of searching for dark matter in our Galaxy by identifying microlensing events toward the

---

galactic bulge (Udalski et al. 1992, 1994a). At times the Bulge is unobservable we conduct other long-term photometric programs. A complete list of side-projects attempted by the OGLE team can be found in Paczyński et al. (1995). In particular, we monitored globular clusters NGC 104 (=47 Tuc) and NGC 5139 (=$\omega$ Cen) in a search for variable stars of various types. Of primary interest was detection of detached eclipsing binaries. Such binaries can potentially provide a vital information about masses of stars in globular clusters. We expected also to detect some contact binaries and pulsating variables of SX Phe type.

It has been known for a long time that binaries play an important role in the dynamical evolution of globular clusters. Different types of binary stars are known to occur in globular clusters (see the detailed review by Hut et al. 1992). Surprisingly very few globular clusters were searched thoroughly for eclipsing binaries. So far positive results were published for NGC 5466 (Mateo et al. 1990), NGC 4371 (Kaluzny & Krzemiński 1993) and M71 (Yan & Mateo 1995; Hodder et al. 1992). Niss et al. (1978) discovered several faint variables in $\omega$ Cen through visual inspection of photographic plates. One of these star, known as NJL-5, turned out to be an eclipsing binary. That was in fact the first variable of this type discovered in any globular cluster. The complete light curve of NJL-5 was published by Jensen & Jorgensen (1985). Recently Gilliland et al. (1995) reported discovery of two eclipsing binaries in the core of 47 Tuc in the data obtained with the HST.

In this paper we report discovery of 11 eclipsing binaries and 7 SX Phe stars in the field of $\omega$ Cen. In addition we obtained light curves for 3 previously known variables - one eclipsing binary, one SX Phe star and one Mira-type star. Photometry of RR Lyr variables in this cluster as well as results for 47 Tuc and 3 other fields in $\omega$ Cen will be published elsewhere.

## 2. Observations and data reduction

The OGLE project is conducted using the 1-m Swope telescope at Las Campanas Observatory which is operated by Carnegie Institution of Washington. A single $2048 \times 2048$ pixels Loral CCD chip, giving the scale of 0.435 arcsec/pixel is used as the detector. The initial processing of the raw frames is done automatically in near-real time. Details of the standard OGLE processing techniques are described by Udalski et al. (1992).

This paper is based in the data obtained during 1993 and 1994 observing seasons. Each season the cluster was monitored during about 3 months. Detailed logs of observations can be found in Udalski et al. (1994, 1995). In 1993 we monitored three fields. Fields 5139A and 5139C were centered about 13 arcmin North and about 13 arcmin South of the cluster center, respectively. Field 5139B covered central part of the cluster. In 1994 we monitored mostly field 5139BC. This field is offset by about 0.5 arcmin South relatively to the field 5139B − such a shift was applied in order to include in the central field a newly discovered detached eclipsing binary which was located in the overlapping area of fields 5139B and 5139C. Field 5139C was also monitored for a few nights at the beginning of the 1994 observing season. Fields 5139A and 5139C overlap by about 1.5 arcmin with fields 5139B and 5139BC. The equatorial coordinates of centers of four observed fields are given in Table 1. Most of the monitoring was performed through the Johnson V filter. Only a few exposures in the Kron-Cousins I band were obtained for fields monitored during the 1993 season. A substantial number of exposures in the I band were obtained for fields 5139C and 5139BC during the 1994 season. Table 2 gives total numbers of frames reduced for each of surveyed fields. For fields 5139A, 5139BC and 5139C and for the V filter the exposure times ranged from 420 sec to 600 sec, with 420 sec being the most frequent value. Most of the V frames of field 5139B were taken with the exposure time set to 240 sec. For majority of the analyzed frames seeing was better than 1.6 arcsec. The DoPHOT photometry program (Schechter et al. 1993) was used to derive profile-fitting photometry. We used DoPHOT in the fixed-position mode. The stellar positions were provided from a list obtained by reduction of "template" images. Separate template list were used for V and I observations. For fields 5139A and 5139C individual frames of particularly good quality served as templates. In case of field 5139BC a template image for the V filter was constructed by averaging 3 individual frames. To cope with positional changes of the point spread function each analyzed frame was divided into a $4 \times 4$ grid of overlapping sub-frames. The point spread function showed only very small variability in these $540 \times 540$ pixel sub-images. Photometry derived for the "template" sub-frames was transformed to the common instrumental system by application of additive corrections. These corrections were equal to the aperture corrections derived for each sub-frame. Subsequently an instrumental photometry derived for a given sub-frame of a given frame was tied to the common instrumental system of the "template" image. Finally the data base containing photometry from all reduced frames was constructed. Relatively poor measurements were flagged at this point. A given measurement was considered to be "poor" if the formal error of photometry was 2.5 times or more larger than the average error of photometry for the nearby stars of comparable magnitudes. The actual procedure of constructing the data base was similar to that described by Szymański & Udalski (1993). Separate data bases were prepared for objects detected on the "template" images and for additional "non-template" objects measured by Dophot on the individual frames. Data bases for the V-band observations were constructed for all four monitored fields. Data bases for the I-band were constructed for fields 5139C and 5139BC, only. Only single frames in the I-band were reduced for fields 5139A and 5139B.

Table 1: Equatorial coordinates for centers of $\omega$ Cen fields monitored for variables by OGLE in 1993 and 1994 seasons.

| Field | $\alpha_{2000}$ | $\delta_{2000}$ |
|---|---|---|
| 5139A | 13:26:38.6 | -47:15:40 |
| 5139B | 13:26:43.8 | -47:28:52 |
| 5139BC | 13:26:42.5 | -47:29:27 |
| 5139C | 13 26 50.8 | -47 42 14 |

The data were calibrated using the following procedure. First we obtained VI photometry for the field located east of the cluster center (field 5139D in Udalski et al. (1995)). This field includes several secondary standards calibrated by Walker (1994). Instrumental photometry for field 5139D was transformed to the standard system using relations:

$$v = const + V - 0.028 \times (V - I) \qquad (1)$$

$$v - i = const + 0.952 \times (V - I) \qquad (2)$$

where a lower case letters correspond to the instrumental magnitudes. The color terms of this transformation were determined using photometry for several Landolt (1993) fields. The zero points were derived using local standards from Walker (1994). Subsequently stars from the overlapping part of fields 5139D and 5139BC were used to establish zero points of VI photometry for stars from field 5139BC. Finally stars from the overlapping parts of field 5139BC and three remaining fields were used to determine zero points of VI photometry for these fields.

The instrumental $v$ and $i$ magnitudes for stars included in data bases were transformed to the VI system using simplified transformation in the form $V = v + const$ and $I = i + const$. To determine value of offsets an average values of $(V - v)$ and $(I - i)$ were calculated for stars with $0.7 < V - I < 1.1$

The color term of the V transformation equals 0.028 (see eq. 1). Hence, neglection of this term leads to some systematic errors of V magnitudes which do not exceed 0.028 mag for stars with $0 < V - I < 2.0$. Colors of variables stars discussed later in this paper range from $V - I \approx 0.3$ to $V - I \approx 1.1$.

## 3. Selection of variable stars

A search for variables was conducted using the data bases for the V filter. Two methods were employed. The first method makes use of a $\chi^2$ statistic. We dropped from the analysis the frames suffering from poor seeing (the limiting value was set to 1.65 arcsec for fields 5139A and 5139C, and to 1.5 arcsec for fields 5139B and 5139BC) and frames with the very high level of sky background. The actual number of frames retained for fields 5139A, 5139B, 5139BC and 5139C, was 144, 100, 204 and 190, respectively. Measurements flagged as "poor" were rejected during extraction of the light curve for a given object. For every star which was measured on at least 100 frames for a given data set, we calculated a $\chi^2$ statistic. To take into account a noise introduced by the procedure of transforming photometry from different frames to the common photometric system and to compensate for the residual effects of variable PSF, we raised formal errors of photometry returned by DoPHOT by 0.02 mag. Objects with $P(\chi^2) < 10^{-4}$ (eg. Press et al. 1986) were considered candidate variables and their light curves were tested for variability with periods in the range from 0.03 to 20 days. To determine the most probable periods we used an *aov* statistic (Schwarzenberg-Czerny 1989, 1991). This statistic allows − in particular − reliable determination of periods for variables with non-sinusoidal light curves (eg. eclipsing variables of EA-type). Phased light curves of suspected variables were subsequently inspected visually and objects with regular light curves were selected. In Fig. 1 we present a plot of *rms* deviation versus the average V magnitude for the light curves of stars from field 5139A examined for variability. The analogous plot for field 5139C looks very much the same and therefore it is not shown. For fields 5139A and 5139C the search for variable stars was conducted down to $V = 19.25$. For fields 5139B and 5139BC the quality of photometry degrades significantly with decreasing distance from the clusters center. Stars from these two fields were divided in two subgroups. For stars located inside central area of size $1000 \times 1000$ pixel$^2$ the search for variables was conducted down to $V = 18.25$ and $V = 17.75$ for fields 5139B and 5139BC, respectively. The outer regions of fields 5139B and 5139BC were searched for variables down to $V = 18.75$. In Figs. 2-3 we plot *rms* vs. $< V >$ for the inner and outer parts of field 5139BC, respectively. Tables 3-5 give some information about the numbers of stars analyzed for variability and numbers of stars which were selected as suspected variables. The total number of stars contained in data bases ranged from about 1.1E5 for fields 5139B and 5139BC to about 3.5E4 for fields 5139A and 5139C. Most of these stars were faint objects with $V > 19$ showing noisy and poorly sampled light curves. Out of about 5300 candidate variables with $< V >> 14.75$ which were selected in all 4 fields with a $\chi^2$ test, only 23 could be classified with confidence as periodic variables. Almost all variables which were included in more than one data base were detected independently in all relevant data sets. The only exceptions were SX Phe variable OGLEGC-2 which was missed in field 5139BC, and an eclipsing binary OGLEGC-15 which was missed in data field 5139B.

Table 2: Number of frames reduced for the $\omega$ Cen fields

| Field | Filter V | Filter I |
|-------|----------|----------|
| 5139A | 238 | 1 |
| 5139B | 152 | 1 |
| 5139BC | 227 | 67 |
| 5139C | 292 | 35 |

The $\chi^2$ statistic allows reliable detection of variables showing sine-like light curves. It is relatively less suitable for detection of detached eclipsing binaries whose light curves are flat for most of the time and show only relatively short episodes of decreased brightness. Therefore the light curves extracted from data bases were scanned with a filter designed to detect probable eclipsing events. A candidate eclipse event was defined as a set of three consecutive points on a light curve which fulfill condition:

$$V + 3 \times \sigma > M(V) + 0.15 \qquad (3)$$

where $\sigma$ is an error of a given measurement and M(V) is a median value of magnitude for a given stars. An additional constrain was that such three observations marking possible eclipse were obtained on the same night. Stars whose light curves exhibited 2 or more possible eclipse events are selected as candidate variables. Table 6 provides numbers of candidates selected in each of four analyzed fields. Comparison of data from Table 6 and Tables 3-5 shows that of the two used methods the first one – based on a $\chi^2$ test – leads to significantly higher frequency of false alarms. All eclipsing binaries discovered using an $\chi^2$ test were detected also by filtering light curves with an "eclipse filter". No new variables were however discovered using the second method of search. Two of the low amplitude variables with sine-like light light curves were discovered only using the $\chi^2$ test.

## 4. Basic properties of variables

The rectangular and equatorial coordinates of 23 periodic variables identified in our data are listed in Table 7. The rectangular coordinates correspond to positions of variables on the V-band "template" images which were submitted to the editors of A&A (see Appendix A). These images allow easy identification of all objects listed in Table 7. The name of field in which a given variable can be identified is given in the 6th column . All frames collected by the OGLE team were deposited at the NASA NSS Data Center [2]. Frames mr4534, mr6489-91 (average of three individual frames), mr4535 and mr4704 were used as templates for fields 5139A, 5139BC, 5139B and 5139C, respectively. For fields 5139A and 5139C transformation from rectangular to equatorial coordinates was derived based on positions of stars from the Guide Star Catalogue (Lasker et al. 1988). We identified 101 and 95 GSC stars for fields 5139A and 5139C, respectively. The adopted plate solutions reproduce equatorial coordinates of GSC stars with residuals rarely exceeding 0.5 arcsec. Transformations for fields 5139BC and 5139B was derived based on positions of 69 RR Lyr variables. Equatorial coordinates for RR Lyr stars in $\omega$ Cen were kindly provided by Dr. Nicolai Samus (Shokin, Evstigneeva and Samus, in preparation). A few GSC stars were identified in the outer parts of fields 5139B and 5139BC. The adopted plate solution reproduces equatorial coordinates of these stars with residuals not exceeding 0.8 arcsec.

Our sample of variables consist of 9 SX Phe stars (variables OGLEGC1-9), 12 binaries (variables OGLEGCC10-21), one object of uncertain classification (OGLEGC-22) and one Mira (OGLEGC-23). Variability of OGLEGC-5 and OGLEGC-19 was first discovered by Niss et al. (1978) and these stars are known as NJL79 and NJL5, respectively. Photometric studies of NJL5 and NJL79 were published by Jensen & Jorgensen (1985) and by Jorgensen & Hansen (1984), respectively. OGLEGC-23 is a previously known Mira-type variable listed as V2 in Hoag (1973) catalogue. This star was near its minimum light during the 1993 observing season It was sufficiently faint at that time to be measured on most of 4 minute long exposures obtained for field 5139B. On the contrary V2 was too bright during the 1994 observing season to be measured on images of field 5139BC (note also longer exposure times for the field 5139BC as compared with the field 5139B).

---

[2]The OGLE data (FITS images) are accessible for astronomical community from the NASA NSS Data Center. Send e-mail to: archives@nssdc.gfc.nasa.gov with the subject line: REQUEST OGLE ALL and put requested frame numbers (in the form MR00NNNN where NNNN stands for frame number according to OGLE notation), one per line, in the body of the message. Requested frames will be available using an "anonymous ftp" service from nssdc.gfc.nasa.gov host in location shown in the return e-mail message from archives@nssdc.gsfc.nasa.gov

Table 8 lists basic characteristic of the light curves of 9 SX Phe star identified in our survey. The mean V magnitudes were calculated by numerically integrating the phased light curves after converting them into intensity scale. Photometric data for the remaining variables are given in Table 9. Phased light curves for stars OGLEGC1-22 are shown in Figs. 4-6. Figure 7 shows the light curve of OGLEGC-23 obtained during the 1993 season.

Figure 8 shows location of variables 1-22 on the cluster color-magnitude diagram (CMD). For SX Phe stars marked positions correspond to the intensity-averaged magnitudes. For the remaining variables we marked positions corresponding to magnitude at maximum light. The main sequence and red giant branch are marked with double lines in Fig. 8. The large width of these sequences is due to the well known spread of metallicity exhibited by stars in $\omega$ Cen (Woolley et al. 1966). All SX Phe stars as well as all contact binaries and two non-contact binaries (OGLEGC-14 and OGLEGC-18) are located on the cluster CMD among candidate blue stragglers. An eclipsing binary OGLEGC-16 is located right at the clusters turnoff while two detached binaries OGLEGC-17 and OGLEGC-15 occupy positions on the subgiant branch.

Light curves of OGLEGC-14, OGLEGC-15 and OGLEGC-17 are flat outside eclipses what indicates that these binaries are detached systems. Each of these stars exhibits two minima of similar depth what implies similar effective temperatures of their components. By combining radial velocity curves with photometry one would be able to determine absolute parameters for components of OGLEGC-14, OGLEGC-15 and OGLEGC-17. OGLEGC-14 is a blue straggler and most probably its components exchanged mass in the past. Relatively long periods of OGLEGC-15 and OGLEGC-17 connected with their positions on the cluster CMD indicate that these systems evolved most probably without mass exchange between their components. Hence, determination of parameters for these two binaries can provide us with a direct measure of masses for the turnoff stars in $\omega$ Cen. Moreover, all three detached systems listed above can be used for determination of a distance to $\omega$ Cen. For eclipsing binaries the fractional accuracy of distance determination is equal to the fractional accuracy of the determination of radial velocity amplitudes, $K_1$ and $K_2$, as the distance is proportional to stellar diameters, which in turn are proportional to the $K$ values. Basing on a spectra collected with large telescopes and using a new "two-dimensional" cross-correlation technique TODOCOR (Zucker & Mazeh 1994; Metcalfe et al. 1995) one may expect to determine radial velocity amplitudes for components of discussed detached binaries with an accuracy better than 1%.

Variable OGLEGC-22 is located on the giant branch in the cluster CMD. The light curve of this stars is unstable (see Fig. 6) what excludes its classification as an ellipsoidal variable. Most probably OGLEGC-22 belongs to so called "spotted variables". It may be either an RS CVn-type star or an FK Com-type star.

### 4.1. Cluster membership of the variables

The $\omega$ Cen cluster is located at an intermediate galactic latitude of $b = +15$ deg. Therefore, one cannot assume a-priori that all variables listed in Table 7 are cluster members. Figure 9 shows the period versus absolute magnitude diagram for 9 SX Phe stars identified in our survey. The standard relations for the F-mode pulsators (solid line) and the H-mode pulsators (dashed line) and for [Fe/H] = −2.2 (upper line) and [Fe/H] = −0.7 (lower line) are also shown. The calibration $P - L - $[Fe/H] was taken from Nemec et al. (1994). We adopted for the cluster an apparent distance modulus $(m - M)_V = 13.86$ while calculating absolute magnitudes of SX Phe stars. The assumed range of metallicities is based on results published by Brown and Wallerstein (1993) and Vanture et al. (1994). One can see that observed luminosities of SX Phe stars observed in the field of $\omega$ Cen are consistent with the hypothesis that all of them are members of the cluster.

We have applied the absolute brightness calibration established by Rucinski (1995) to calculate $M_V$ for newly discovered contact binaries. Rucinski's calibration gives $M_V$ as a function of period, unreddened color $(V - I)_0$ and metallicity:

$$M_V^{cal} = -4.43 log(P) + 3.63(V - I)_0 - 0.31 - 0.12 \times \text{[Fe/H]}. \qquad (4)$$

We adopted for all systems [Fe/H] = −1.6 what is the mean metallicity for the cluster stars. Figure 10 shows the period versus an apparent distance moduli diagram for contact binaries identified in the surveyed area of $\omega$ Cen. An apparent distance modulus was calculated for each system as a difference between its V magnitude and $M_V^{cal}$. An apparent distance modulus for $\omega$ Cen is estimated at $(m - M)_V = 13.86$ (Nemec et al. 1994). We conclude based on Fig. 10 that all newly discovered contact binaries are likely members of the cluster.

### 5. The color-magnitude diagrams

In Fig. 11 we show the $V$ vs. $V - I$ CMDs of three surveyed fields. For each field the photometry is based on a single pair of V&I exposures – the V-filter "templates" (see 1'st paragraph of section 4) were supplemented by the deepest available exposures in the I-band. We obtained VI photometry for a more than 1E5 stars from all 3 fields. The presented CMDs were cleaned from stars of relatively poor photometry. It is known that $\omega$ Cen possesses an extended horizontal branch (Da Costa et al. 1986; Bailyn et al. 1992). Our data show that a sequence of blue stars extends down to $V \approx 19$. Moreover, the data for fields 5139A and 5139C indicate that the sequence of faint blue stars has a gap approximately at $17.4 < V < 18.0$.

Another feature which can easily be noted on CMDs for fields 5139A and 5139C is a scattered sequence of stars located to the blue of the subgiant branch of the cluster. The distribution of colors for these stars shows a cut-off at $(V - I) \approx 0.70$. This scattered sequence corresponds most probably to the foreground stars from the thick disk.

A substantial number of blue straggler candidates can be identified in CMD's for fields 5139A and 5139C. The accuracy of photometry for stars from field 5139BC is relatively low due increasing crowding as we move toward the cluster center. In Fig. 12 we show CMD for these stars from the field 5139BC whose photometry fulfills conditions $\sigma_V \leq 0.025$ and $\sigma_{rmBV} \leq 0.035$. Several candidates for blue stragglers are visible in this figure. In addition to photometry presented in Fig. 11 we constructed another CMDs for the field 5139BC. The CMD obtained by averaging photometry extracted from long and short exposures is show in Fig. 12a. Fig. 12b presents CMD based on frames mr6492 and mr6822 while Fig. 12c shows CMD based on frames mr6492 and mr6821. Some information about frames used to construct CMDs presented in Fig. 12 are given in Table 10. Again presented CMDs were cleaned from stars of relatively poor photometry. An onset of the asymptotic giant branch at $V \approx 14.0$ can be noted in Figs. 12a-c.

Stellar models predict that the luminosity function (LF) of the red giants in globular clusters should show a small "bump" located close to the luminosity level defined by the horizontal branch. In fact such "bumps" have been detected for several clusters (eg. Fusi Pecci et al. 1990). We used data shown if Fig. 12a to derive the observed LF for red giants in $\omega$ Cen. This LF is presented in two forms in Figs. 13 and 14. The LF shown in Fig. 13 shows a break in the slope at $V \approx 14.75$. Following Fusi Pecci et al. (1990) we may assume that this value corresponds to location of the faint edge of the "bump". Fusi Pecci et al. (1990) defined parameter $\Delta V_{bump}^{HB}$, measuring the difference in magnitudes between the horizontal branch and the faint edge of the "bump". This parameter correlates with the cluster metallicity and is negative for clusters with $[Fe/H] < -1.3$. For $\omega$ Cen we obtained $\Delta V_{bump}^{HB} = 14.75 - 14.52 = 0.23$. This is in apparent conflict with the average metallicity of the cluster which is $[Fe/H] = -1.6$. However, examination of Fig. 14 reveals that the "bump" observed on the LF of red giants in $\omega$ Cen is atypically wide (compare with Fig. 4 in Fusi Pecci et al. 1990) and extends from $V \approx 14.30$ to $V \approx 14.75$. This large width can be interpereted as due to a significant spread of metallicity observed for stars in $\omega$ Cen (Wooley et al. 1966).

All photometry presented in this section was submitted in tabular form to the editors of A&A and is available in electronic form to all interested readers (see Appendix B).

## 6. Summary

An extended survey of the central part of $\omega$ Cen lead to discovery of 20 new periodic non-RR Lyr variables. We obtained also light curves of 2 previously known variables. Our sample of variables includes 9 SX Phe stars, 7 contact binaries, 5 non-contact eclipsing binaries and one likely spotted variable. All identified SX Phe stars and all contact binaries occupy positions among blue stragglers in the cluster CMD. Three non-contact binaries are located among turnoff stars. Of particular interest are detached systems OGLEGC15 and OGLEGC17 which can potentially provide information about massess of turnoff stars in $\omega$ Cen.

In the next paper from this series we shall present results for another 3 $\omega$ Cen fields which were monitored by the OGLE team during the 1995 observing season. A separate paper will be devoted to RR Lyr stars in $\omega$ Cen.

## 7. Appendix A

In addition to certain periodic variables listed in Table 7, we identified in the field 5139B another two likely periodic variables. Rectangular and equatorial coordinates of these stars are given in Table 11. Images of both stars are badly crowded and their V-band photometry is relatively poor. Moreover, we failed to determine $V - I$ colors for these stars. OGLEGC24 shows variability with a period $P \approx 0.0533$ day and its magnitude varies from $V_{\min} \approx 17.3$ to $V_{\max} \approx 17.1$. It belongs probably to SX Phe-type stars. One likely eclipse event (about 0.30 mag deep) was observed for OGLEGC25. An out-of-eclipse magnitude of this star is $V \approx 17.0$.

## 8. Appendix B

Tables containing light curves of all variables discussed in this paper as well as tables with VI photometry for more then 1E5 stars from the surveyed fields are published by A&A at the centre de Données de Strasbourg, where they are available in electronic form: See the Editorial in A&A 1993, Vol. 280, page E1. Equatorial coordinates are given for all stars included in these tables. We have also submitted the V-filter "template" images of four analyzed fields. These images can be used for identification of all variables discussed in this paper as well as for identification of all stars for which we provide VI photometry.

This project was supported by NSF grants AST 92-16494 and AST-9313620 to Bohdan Paczynski and AST 92-16830 to George Preston. MK, MS and AU were supported by the Polish KBN grant 2P03D02908 to Andrzej Udalski. JK was supported by the Polish KBN grant 2P03D00808. JK wishes to thank Tomek Plewa for countless tips about C-shell scripts. We are indepted to Dr Nicolai Samus for providing us with equatorial coordinates of RR Lyr stars in $\omega$ Cen.

Table 3: Basic statistical data about stars from fields 5139A and 5139C examined for variability using an $\chi^2$ test. The data are given for bins 0.5 mag wide. Columns 2 and 5 give *median* value of *rms* for a given bin. Columns 3 and 6 give number of stars examined for variability while columns 4 and 7 give number of suspected variables selected using the $\chi^2$ test.

| V | A | | | C | | |
| | $<rms>$ | N | VAR | $<rms>$ | N | VAR |
|---|---|---|---|---|---|---|
| 14.5 | 0.011 | 100 | 18 | 0.013 | 86 | 5 |
| 15.0 | 0.009 | 168 | 9 | 0.013 | 161 | 2 |
| 15.5 | 0.009 | 153 | 6 | 0.013 | 139 | 3 |
| 16.0 | 0.011 | 202 | 14 | 0.013 | 132 | 2 |
| 16.5 | 0.013 | 276 | 6 | 0.015 | 244 | 4 |
| 17.0 | 0.017 | 401 | 18 | 0.017 | 339 | 25 |
| 17.5 | 0.022 | 877 | 61 | 0.022 | 717 | 41 |
| 18.0 | 0.027 | 1946 | 141 | 0.028 | 1633 | 137 |
| 18.5 | 0.038 | 3255 | 338 | 0.039 | 2813 | 263 |
| 19.0 | 0.055 | 4500 | 501 | 0.056 | 4094 | 500 |

Table 4: Same as Table 3 but for inner parts of fields 5139B and 5139BC.

| V | B $< rms >$ | N | VAR | BC $< rms >$ | N | VAR |
|---|---|---|---|---|---|---|
| 14.5 | 0.012 | 86 | 6 | 0.027 | 206 | 32 |
| 15.0 | 0.012 | 105 | 2 | 0.025 | 359 | 35 |
| 15.5 | 0.014 | 105 | 2 | 0.026 | 327 | 31 |
| 16.0 | 0.019 | 118 | 2 | 0.031 | 392 | 62 |
| 16.5 | 0.028 | 194 | 14 | 0.041 | 577 | 134 |
| 17.0 | 0.037 | 285 | 25 | 0.059 | 1199 | 403 |
| 17.5 | 0.046 | 558 | 51 | 0.077 | 2555 | 1039 |
| 18.0 | 0.064 | 975 | 143 | | | |

Table 5: Same as Table 3 but for outer parts of fields 5139B and 5139BC

| V | B | | | BC | | |
| | $< rms >$ | N | VAR | $< rms >$ | N | VAR |
|---|---|---|---|---|---|---|
| 14.5 | 0.009 | 84 | 12 | 0.017 | 87 | 12 |
| 15.0 | 0.010 | 123 | 4 | 0.014 | 128 | 6 |
| 15.5 | 0.011 | 114 | 2 | 0.013 | 118 | 2 |
| 16.0 | 0.014 | 144 | 4 | 0.015 | 164 | 2 |
| 16.5 | 0.018 | 236 | 8 | 0.018 | 241 | 6 |
| 17.0 | 0.024 | 334 | 8 | 0.024 | 328 | 10 |
| 17.5 | 0.032 | 874 | 36 | 0.029 | 847 | 53 |
| 18.0 | 0.046 | 2002 | 166 | 0.038 | 1825 | 208 |
| 18.5 | 0.065 | 3282 | 332 | 0.053 | 3042 | 473 |

Table 6: Number of stars selected as potential variables by scanning light curves with a filter designed to pick-up eclipse-like events. Columns 2-5 give results for fields 5139A, 5139C and for the outer parts of fields 5139B and 5139BC.

| V | A $N_{\mathrm{VAR}}$ | C $N_{\mathrm{VAR}}$ | B $N_{\mathrm{VAR}}$ | BC $N_{\mathrm{VAR}}$ |
|---|---|---|---|---|
| 14.5 | 15 | 5 | 9 | 10 |
| 15.0 | 1 | 1 | 1 | 3 |
| 15.5 | 0 | 0 | 0 | 0 |
| 16.0 | 2 | 0 | 1 | 2 |
| 16.5 | 1 | 1 | 2 | 4 |
| 17.0 | 0 | 10 | 1 | 8 |
| 17.5 | 2 | 14 | 1 | 12 |
| 18.0 | 7 | 44 | 4 | 49 |
| 18.5 | 23 | 46 | 6 | 60 |
| 19.0 | 29 | 54 | 7 | 56 |

Table 7: Rectangular and equatorial coordinates for variables identified in $\omega$ Cen. The X and Y coordinates give positions of variables on the template images (see text). The last column gives alternative names for three variables which were previously known.

| Name | X | Y | RA(2000) (h:m:s) | Dec(2000) (deg:':") | Field | Other Name |
|------|------|------|------|------|------|------|
| OGLEGC-1 | 458.02 | 693.52 | 13:26:20.45 | -47:31:59.77 | 5139BC | |
| OGLEGC-2 | 796.44 | 794.47 | 13:26:34.55 | -47:31:03.44 | 5139BC | |
| OGLEGC-3 | 687.99 | 1139.61 | 13:26:28.64 | -47:28:38.03 | 5139BC | |
| OGLEGC-4 | 1606.97 | 1401.50 | 13:27:06.90 | -47:26:10.17 | 5139BC | |
| OGLEGC-5 | 1829.46 | 1603.02 | 13:27:15.62 | -47:24:34.54 | 5139BC | NJL79 |
| OGLEGC-6 | 353.63 | 1019.25 | 13:26:11.04 | -47:15:53.90 | 5139A | |
| OGLEGC-7 | 1048.81 | 1524.93 | 13:26:38.85 | -47:11:51.45 | 5139A | |
| OGLEGC-8 | 1967.81 | 966.84 | 13:27:19.91 | -47:15:21.08 | 5139A | |
| OGLEGC-9 | 1450.04 | 1676.90 | 13:27:07.87 | -47:37:05.26 | 5139C | |
| OGLEGC-10 | 862.94 | 1905.14 | 13:26:33.28 | -47:23:00.11 | 5139BC | |
| OGLEGC-11 | 1485.72 | 42.80 | 13:27:06.86 | -47:36:03.07 | 5139BC | |
| OGLEGC-12 | 1389.98 | 1819.19 | 13:26:56.07 | -47:23:17.58 | 5139BC | |
| OGLEGC-13 | 1691.80 | 1401.10 | 13:27:10.52 | -47:26:07.14 | 5139BC | |
| OGLEGC-14 | 486.71 | 1898.47 | 13:26:17.27 | -47:23:16.98 | 5139BC | |
| OGLEGC-15 | 1033.92 | 89.44 | 13:26:47.34 | -47:35:59.91 | 5139BC | |
| OGLEGC-16 | 1188.07 | 1840.97 | 13:26:47.38 | -47:23:15.73 | 5139BC | |
| OGLEGC-17 | 1983.73 | 1732.20 | 13:27:21.71 | -47:23:32.78 | 5139BC | |
| OGLEGC-18 | 1948.51 | 1967.27 | 13:27:19.31 | -47:21:52.40 | 5139BC | |
| OGLEGC-19 | 1640.67 | 1897.75 | 13:27:02.67 | -47:08:49.10 | 5139A | NJL5 |
| OGLEGC-20 | 1769.64 | 1618.77 | 13:27:21.76 | -47:37:19.14 | 5139C | |
| OGLEGC-21 | 524.39 | 1107.09 | 13:26:30.08 | -47:41:44.32 | 5139C | |
| OGLEGC-22 | 191.66 | 917.77 | 13:26:08.24 | -47:30:32.52 | 5139BC | |
| OGLEGC-23 | 319.91 | 1630.78 | 13:26:12.64 | -47:24:42.14 | 5139B | V2 |

Table 8: Light-curve parameters for SX Phe stars from the field of $\omega$ Cen. $A_V$ is the range of observed variations in the V band. (V-I) is the observed color at the maximum light. The period is given in days.

| Name | Period | $<V>$ | $V$ max | $A_V$ | (V-I) |
|------|--------|-------|---------|-------|-------|
| OGLEGC-1 | 0.0471210 | 17.03 | 16.95 | 0.13 | 0.58 |
| OGLEGC-2 | 0.0481816 | 17.41 | 17.32 | 0.15 | 0.71 |
| OGLEGC-3 | 0.0622868 | 16.65 | 16.21 | 0.63 | 0.65 |
| OGLEGC-4 | 0.0495208 | 16.72 | 16.52 | 0.32 | 0.47 |
| OGLEGC-5 | 0.0654911 | 16.79 | 16.56 | 0.36 | 0.50 |
| OGLEGC-6 | 0.0506528 | 17.20 | 17.09 | 0.19 | 0.54 |
| OGLEGC-7 | 0.0464247 | 17.17 | 17.11 | 0.12 | 0.57 |
| OGLEGC-8 | 0.0417849 | 16.75 | 16.61 | 0.25 | 0.49 |
| OGLEGC-9 | 0.0493750 | 16.95 | 16.71 | 0.40 | 0.41 |

Table 9: Light-curve parameters for variables OGLEGC10-23 from the field of $\omega$ Cen. (V-I) is the observed color at the maximum light. $T_0$ is the time of minimum light. Variable OGLEGC-22 is listed twice – the first entry corresponds to the 1993 season while the second one corresponds to the 1994 season.

| Name | Type | Period (days) | $V_{\max}$ | $V_{\min}$ | $T_0$ HJD 2449000+ | (V-I) |
|---|---|---|---|---|---|---|
| OGLEGC-10 | EW | 0.368712 | 17.31 | 17.65 | 81.7776 | 0.69 |
| OGLEGC-11 | EW | 0.307348 | 17.20 | 17.50 | 81.9880 | 0.52 |
| OGLEGC-12 | EW | 0.275865 | 17.79 | 18.10 | 464.7160 | 0.56 |
| OGLEGC-13 | EW | 0.305525 | 17.12 | 17.44 | 81.7978 | 0.44 |
| OGLEGC-14 | EA | 0.834420 | 16.64 | 16.92 | 81.6456 | 0.31 |
| OGLEGC-15 | EA | 1.496605 | 16.96 | 17.58 | 82.6188 | 1.08 |
| OGLEGC-16 | EA | 0.576233 | 18.15 | 19.38 | 81.9842 | 0.82 |
| OGLEGC-17 | EA | 2.466760 | 17.27 | 17.60 | 82.3608 | 0.89 |
| OGLEGC-18 | EA | 1.376188 | 16.02 | 17.15 | 82.3725 | 0.44 |
| OGLEGC-19 | EW | 0.39823 | 16.22 | 16.51 | 81.8333 | 0.38 |
| OGLEGC-20 | EW | 0.341803 | 16.61 | 16.78 | 81.7196 | 0.44 |
| OGLEGC-21 | EW | 0.249260 | 17.86 | 18.05 | 81.7409 | 0.66 |
| OGLEGC-22 | RS | 22 | 15.115 | 15.232 | 81.6 | — |
| OGLEGC-22 | RS | 22 | 15.075 | 15.182 | 464.5 | 1.11 |
| OGLEGC-23 | M | 236 | — | 15.78 | — | — |

Table 10: List of frames used for construction of CMDs shown in Fig. 12

| Frame | $T_{\exp}$ sec | Filter | FWHM arcsec |
|-------|------|--------|------|
| mr6489 | 420 | V | 0.92 |
| mr6490 | 420 | V | 0.92 |
| mr6491 | 300 | V | 0.92 |
| mr6492 | 60 | V | 0.94 |
| mr6425 | 300 | I | 1.15 |
| mr6821 | 15 | I | 1.06 |
| mr6822 | 60 | I | 1.08 |

Table 11: Rectangular and equatorial coordinates for two suspected variables OGLEGC24-25 identified in the field 5139B.

| Name | X | Y | RA(2000) (h:m:s) | Dec(2000) (deg:':") |
|------|-----|-----|-----|-----|
| OGLEGC-23 | 390.20 | 1247.84 | 13:26:16.91 | -47:27:25.6 |
| OGLEGC-24 | 1812.55 | 524.36 | 13:27:20.22 | -47:31:49.7 |

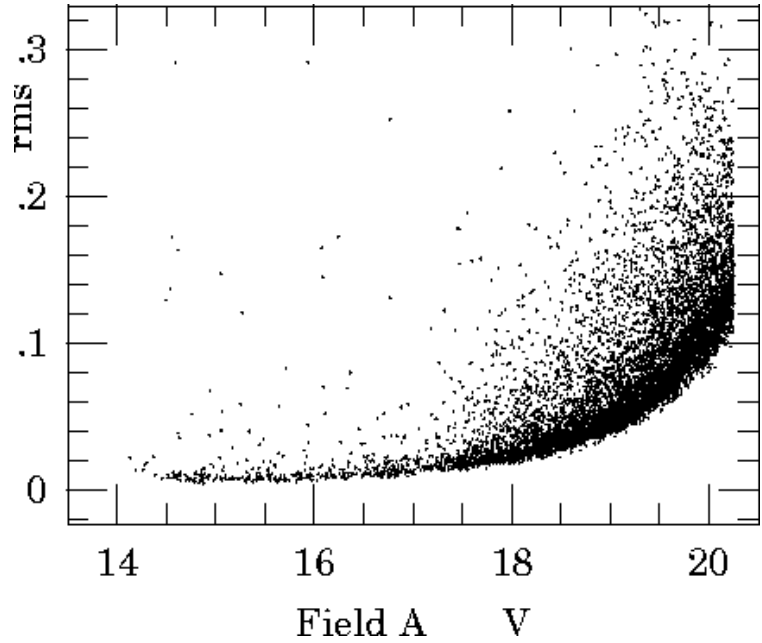

Fig. 1.— Standard deviation versus average V magnitude for stars with at least 100 reliable measurements. This plot corresponds to field 5139A. Only 1/3 of the whole sample of stars examined for variability was plotted for clarity. Stars with $V > 19.25$ are not shown as they were not examined for variability.

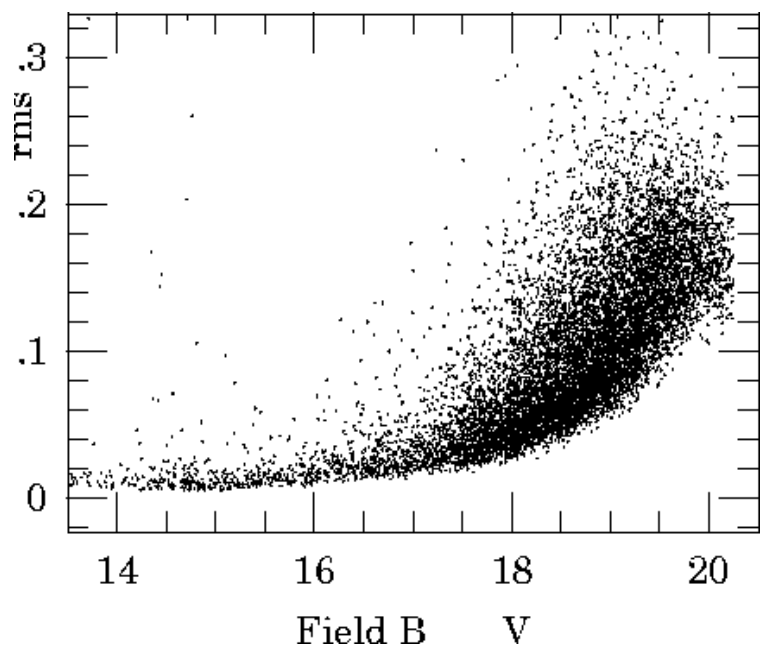

Fig. 2.— Same as Fig. 1 but for the inner part of field 5139BC. Only 1/5 of the whole sample was plotted for clarity.

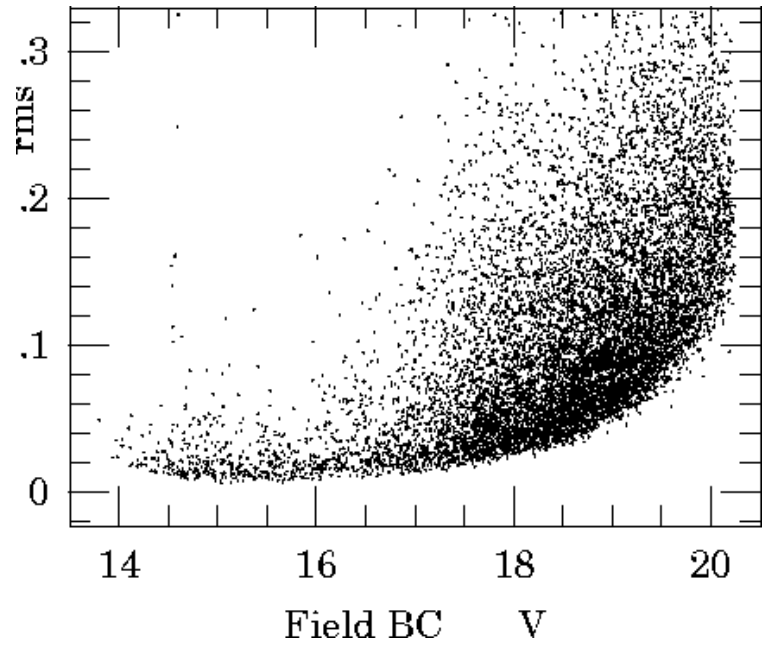

Fig. 3.— Same as Fig. 1 but for the outer part of field 5139BC. Only 1/5 of the whole sample was plotted for clarity.

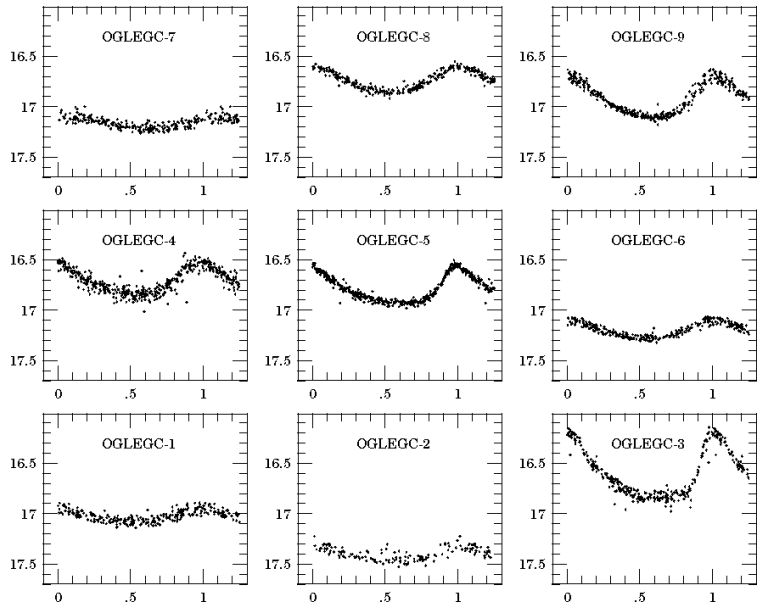

Fig. 4.— Phased V light curves for 9 SX Phe stars identified in $\omega$ Cen. Note the same height of all boxes.

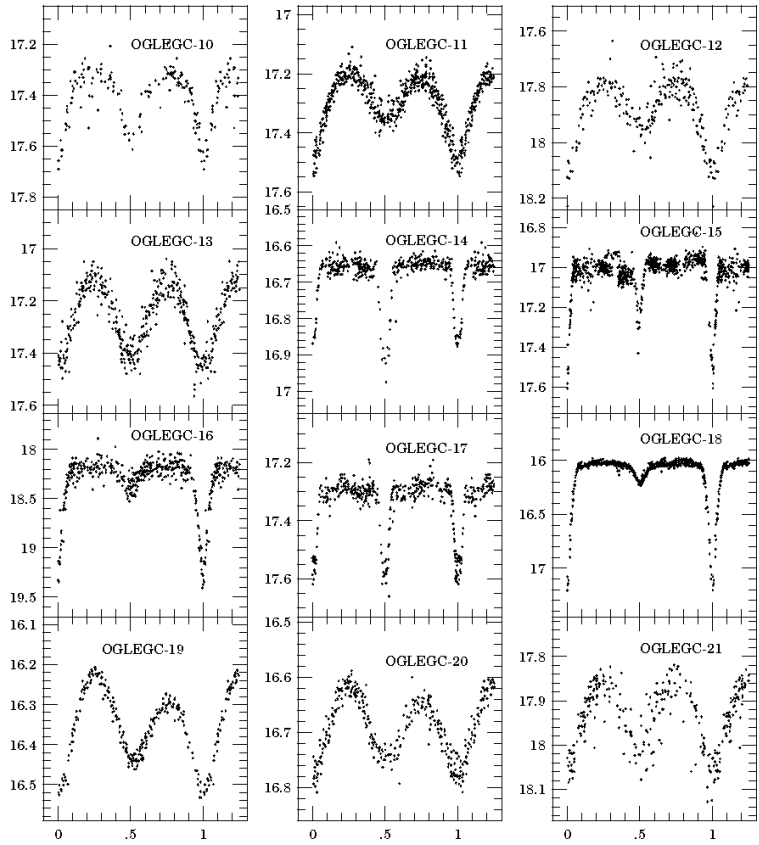

Fig. 5.— Phased V light curves for variables OGLEGC10-21

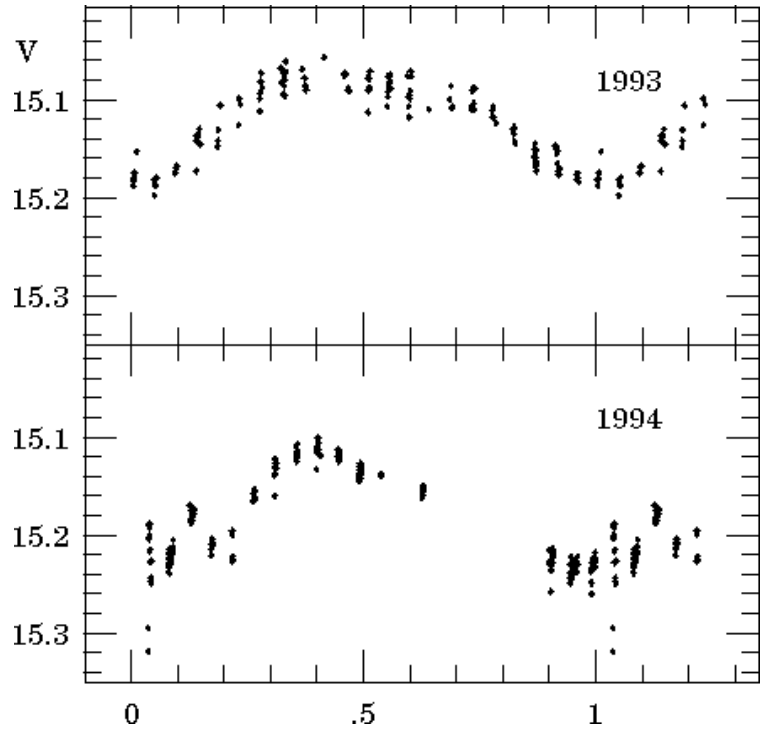

Fig. 6.— Phased V light curves for variable OGLEGC-22. Upper light curve corresponds to 1993 season while the lower one corresponds to 1994 season.

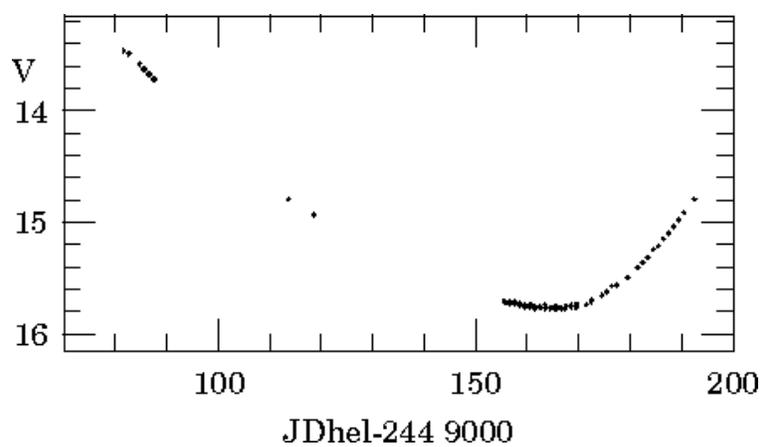

Fig. 7.— The V-band light curve for variable OGLEGC-23.

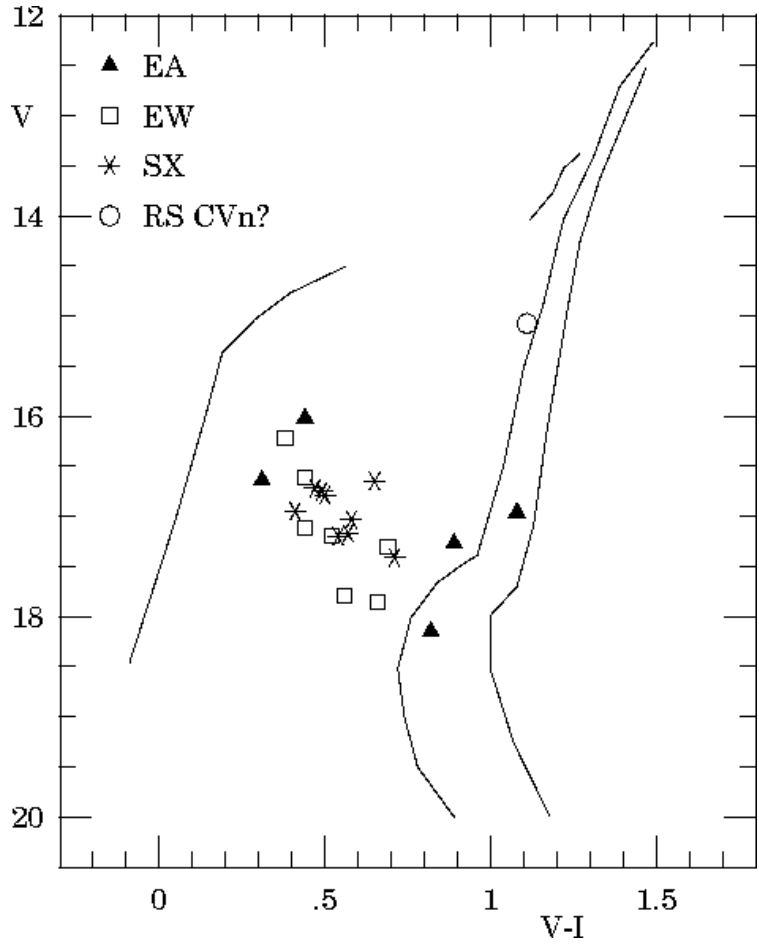

Fig. 8.— The schematic CMD for $\omega$ Cen with positions of variables OGLEGC1-22 marked. The triangles represent contact binaries, the circles SX Phe stars and squares non-contact eclipsing binaries. Probable RS CVn-type variable OGLEGC-22 is marked with an open circle.

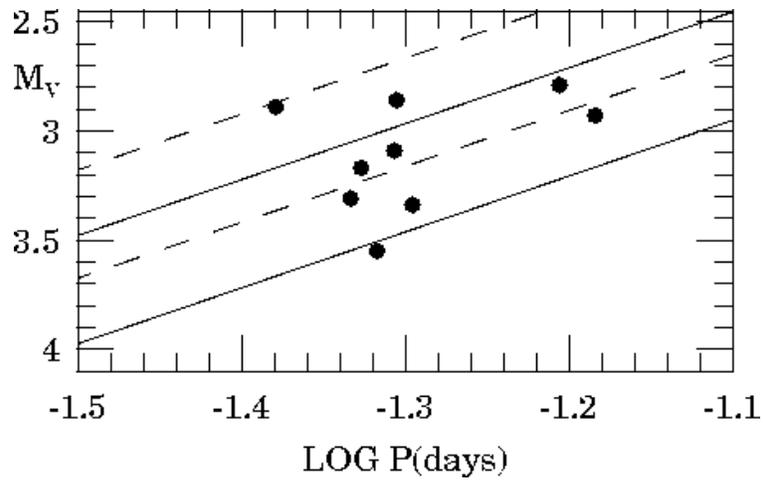

Fig. 9.— Period, absolute magnitude diagram for SX Phe stars from the field of $\omega$ Cen.

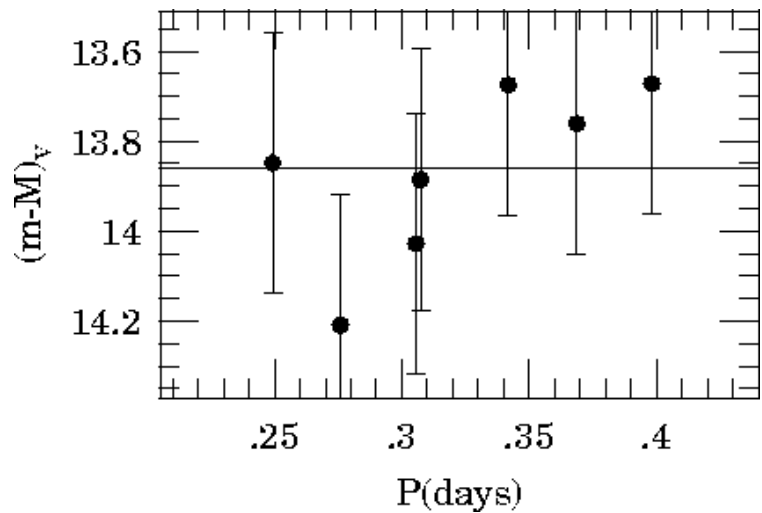

Fig. 10.— Period, apparent distance modulus diagram for contact binaries from the field of $\omega$ Cen. A horizontal line at $(m - M)_V = 13.86$ corresponds to distance modulus of the cluster. Error bars correspond to the formal uncertainty of absolute magnitudes derived using Rucinski's (1995) calibration.

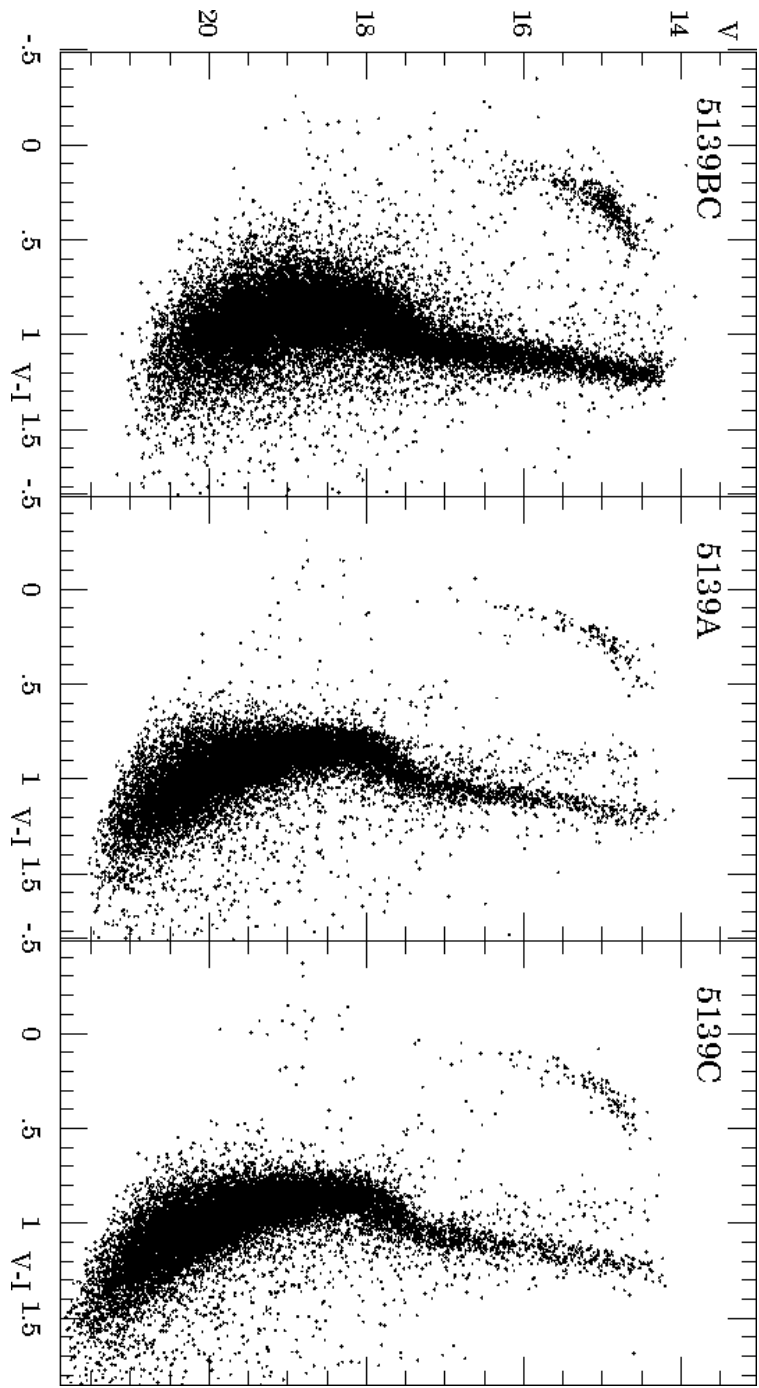

Fig. 11.— The CMDs for fields 5139BC (left), 5139A (center) and 5139C (right).

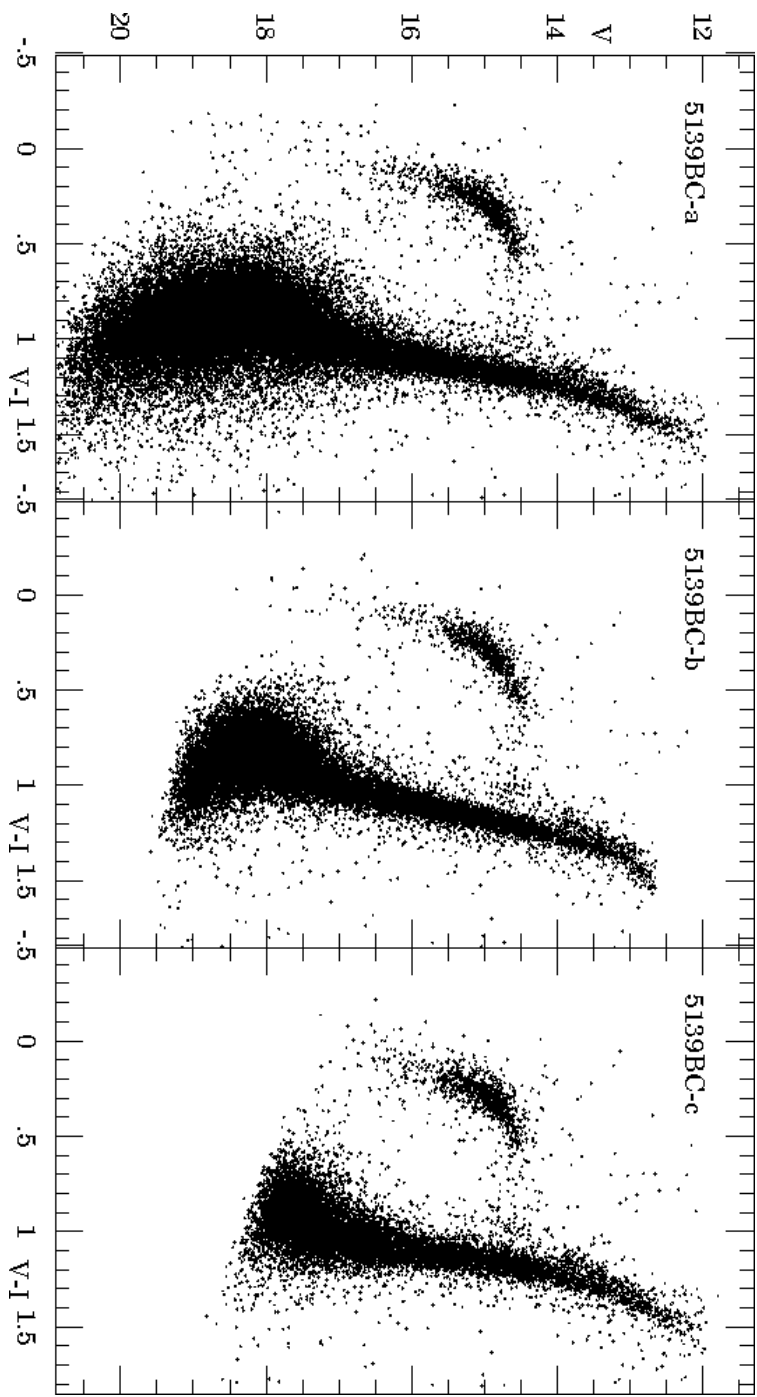

Fig. 12.— The CMD's for field 5139BC (see text for details).

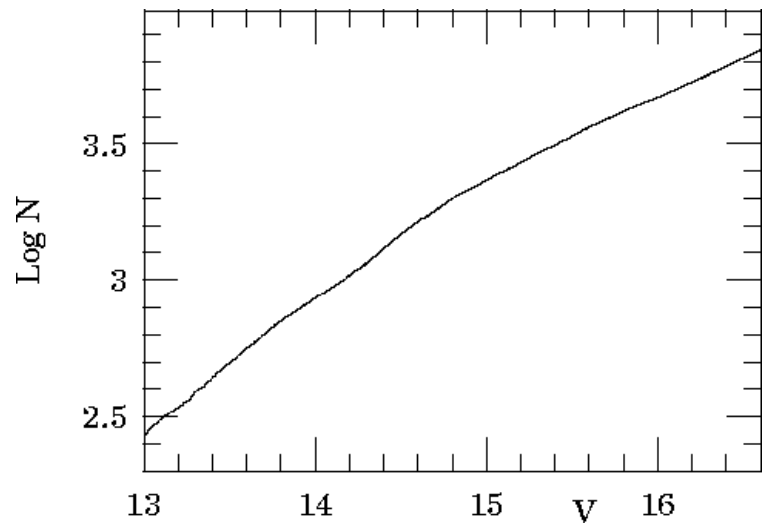

Fig. 13.— The observed itegrated luminosity function for stars from the red giant branch of $\omega$ Cen.

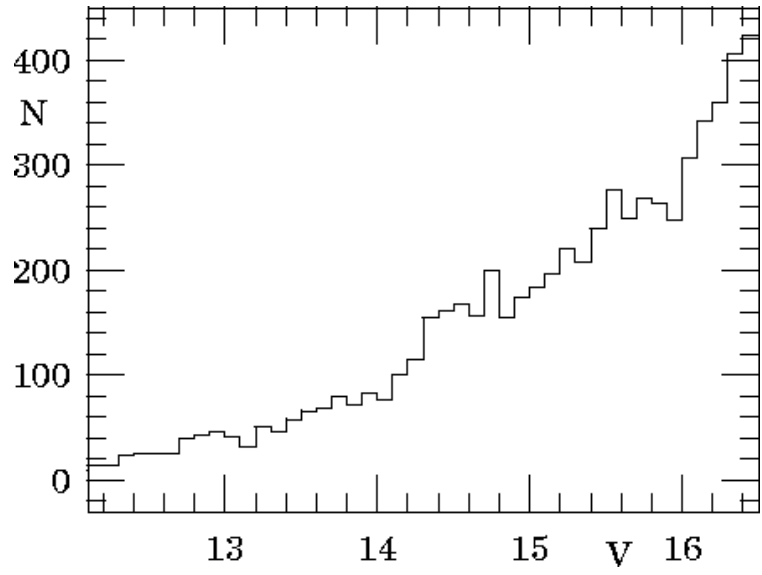

Fig. 14.— The observed differential luminosity function for stars from the red giant branch of ω Cen.